\documentclass{PoS}

\def\beqa{\begin{eqnarray*}}
\def\eeqa{\end{eqnarray*}}

\title{$D^{+}$, $D^{0}$ and $\Lambda_{c}^{+}$ production in deep inelastic scattering at HERA}

\ShortTitle{$D^{+}$, $D^{0}$ and $\Lambda_{c}^{+}$ production in DIS at HERA}

\author{\speaker{Philipp Roloff}$^{ab}$\thanks{On behalf of the ZEUS Collaboration.}\\
        \llap{$^{a}$}Deutsches Elektronen-Synchrotron DESY, Notkestr. 85, D-22607 Hamburg, Germany\\
        \llap{$^{b}$}University of Hamburg, Institute of Experimental Physics, Luruper Chaussee 149, D-22761 Hamburg, Germany\\
        E-mail: \email{philipp.roloff@desy.de}}

\abstract{Several recent measurements of charmed hadron production in deep inelastic scattering at HERA are reviewed. Cross sections for the production of $D$ mesons were measured and compared to NLO QCD predictions. The charm contribution to the inclusive structure function $F_{2}$ was extracted and compared to previous measurements and theoretical predictions. The reconstruction of decays with a neutral strange hadron in the final state allowed the measurement of charm production to be extended into the low transverse momentum region. The fraction of $c$ quarks hadronising into $\Lambda_{c}^{+}$ baryons was extracted using two different decay modes.}

\FullConference{35th International Conference of High Energy Physics - ICHEP2010,\\
		July 22-28, 2010\\
		Paris France}

\begin{document}

\section{Introduction}

Charm production in the deep inelastic scattering (DIS) regime (squared four-momentum exchange at the electron vertex $Q^{2} \gtrsim$ a few GeV$^{2}$) is dominated by the boson-gluon fusion (BGF) process, where the virtual photon interacts with a gluon from the proton. Hence this process is directly sensitive to the gluon content of the proton. The contribution of charm production to the inclusive proton structure function $F_{2}$ can be extracted from double-differential cross sections as a function of the Bj{\o}rken scaling variable, $x$, and $Q^{2}$.

The hadronisation of charm quarks into colourless hadrons is not calculable by perturbative Quantum Chromodynamics (pQCD). Thus properties of the hadronisation process like fragmentation fractions of charmed hadrons need to be extracted experimentally.

\section{Measurement of $D^{+}$ at threshold and $\Lambda_{c}^{+}$ production}

Charm quark production was measured with the ZEUS detector at HERA reconstructing the hadronic decay channels $D^{+} \rightarrow K^{0}_{S}\pi^{+}$, $\Lambda_{c}^{+} \rightarrow pK^{0}_{S}$ and $\Lambda_{c}^{+} \rightarrow \Lambda\pi^{+}$, and their charge conjugates, using an integrated luminosity of $120.4 \pm 2.4$~pb$^{-1}$~\cite{ref:dplus_lambdac_paper}. The presence of a neutral strange hadron in the final state reduced the combinatorial background and extended the measured sensitivity into the low transverse momentum region. The kinematic range of the measurement is given by: $0 < p_{T}(D^{+}, \Lambda_{c}^{+}) < 10$~GeV, $|\eta(D^{+}, \Lambda_{c}^{+})| < 1.6$, $1.5 < Q^{2} < 1000$~GeV$^{2}$ and $0.02 < y < 0.7$.

\begin{figure}[h]
\includegraphics[width=0.5\textwidth]{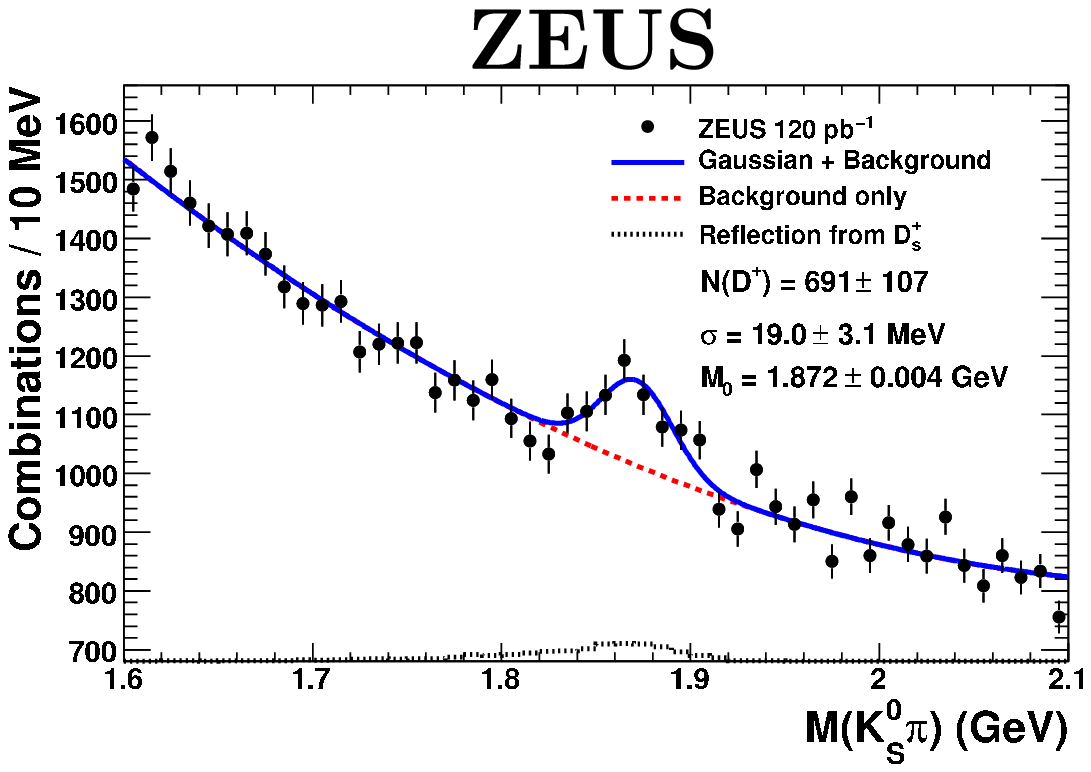}
\includegraphics[width=0.5\textwidth]{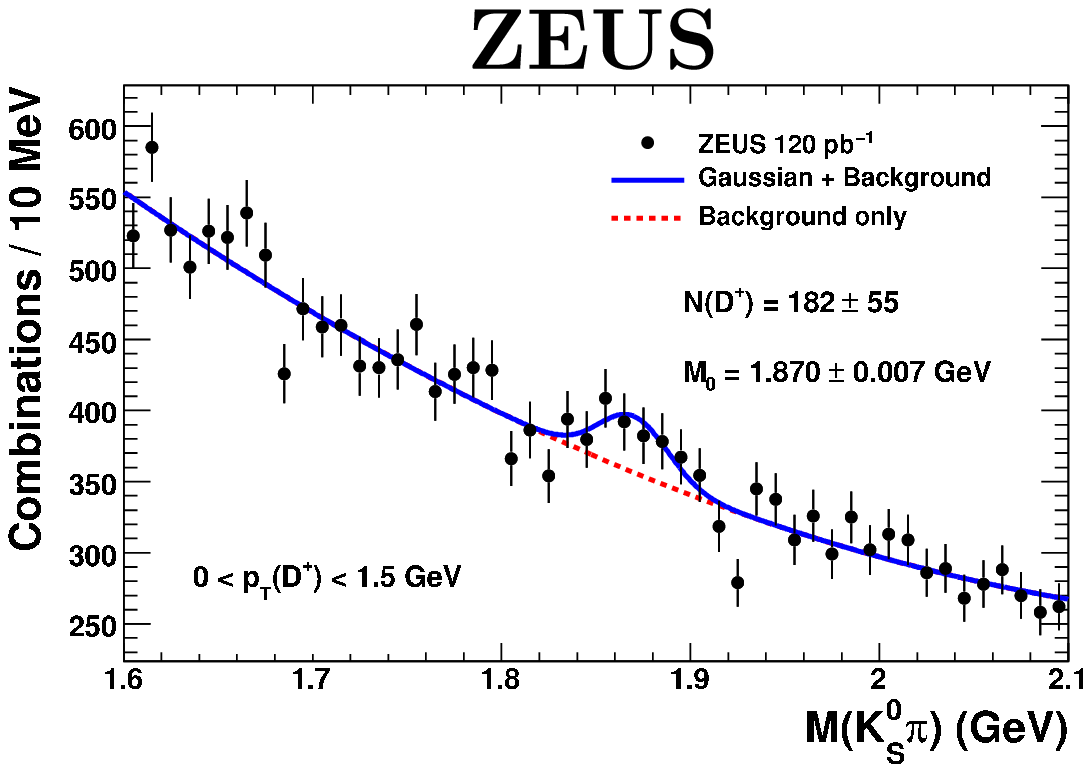}
\caption{The $M(K^{0}_{S}\pi^{+})$ distribution (dots) for $D^{+}$ candidates. The reflection caused by the decay $D_{s}^{+} \rightarrow K^{0}_{S}K^{+}$ has been subtracted. The left plot shows the signal in the full kinematic region while the plot on the right side shows the signal for the restricted range $0 < p_{T}(D^{+}) < 1.5$~GeV.}
\label{fig:dplus_masspeaks_hera1}
\end{figure}

The observed invariant mass spectra for the $D^{+}$ candiates are shown in Fig.~\ref{fig:dplus_masspeaks_hera1} for the full kinematic region of the measurement and for the restricted range $0 < p_{T}(D^{+}) < 1.5$~GeV. Signals at the nominal $D^{+}$ mass are visible in both distributions. This measurement allowed the production of $D$ mesons to be studied for the first time at HERA in the region $p_{T}(D) < 1.5$~GeV.

\begin{figure}[h]
\centerline{\includegraphics[width=0.4\textwidth]{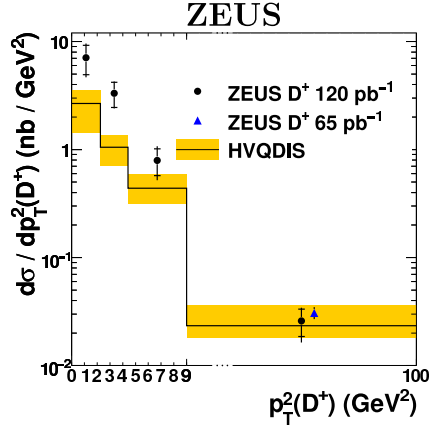}}
\caption{Differential $D^{+}$ cross section as a function of $p_{T}^{2}(D^{+})$ compared to the NLO QCD calculation of HVQDIS. The measured values are shown as dots while the triangle represents a previous ZEUS result. Note the discontinuity in the $X$-axis between $9$ and $100$~GeV$^{2}$.}
\label{fig:dplus_cross_section_pt2_hera1}
\end{figure}

The measured inclusive and differential cross sections are reasonably well described by next-to-leading-order (NLO) QCD predictions from the HVQDIS program~\cite{ref:hvqdis} based on the fixed-flavour-number scheme (FFNS). The FFNS variant of the ZEUS-S NLO QCD fit~\cite{ref:zeus_s} to structure function data was used as the parametrisation of the proton PDFs. The charm quark mass was set to $m_{c} = 1.5$~GeV and the renormalisation and factorisation scales were set to $\mu_{R} = \mu_{F} = \sqrt{Q^{2} + 4m_{c}^{2}}$. As an example, Fig.~\ref{fig:dplus_cross_section_pt2_hera1} shows the differential $D^{+}$ cross section as a function of $p_{T}^{2}(D^{+})$. For $p_{T}^{2}(D^{+}) > 9$~GeV$^{2}$, the measurement is in good agreement with a previous ZEUS result~\cite{ref:d_mesons_dis_zeus_hera1}.

\begin{figure}[h]
\centerline{\includegraphics[width=0.55\textwidth]{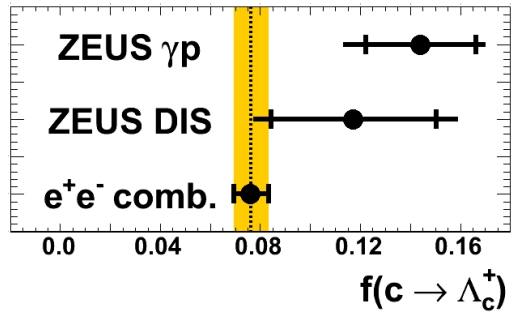}}
\caption{The fraction of $c$ quarks hadronising to a $\Lambda_{c}^{+}$ baryon.}
\label{fig:f_lambdac_hera1}
\end{figure}

Additionally, the fraction of $c$ quarks hadronising into $\Lambda_{c}^{+}$ hadrons, $f(c \rightarrow \Lambda_{c}^{+})$, was extracted from a combination of the results obtained using both $\Lambda_{c}^{+}$ decay channels mentioned above. The following result was obtained:
\beqa
f(c \rightarrow \Lambda_{c}^{+}) = 0.117 \pm 0.033 ({\rm stat.}) ^{+0.026}_{-0.022} ({\rm syst.}) \pm 0.027 ({\rm br.}).
\eeqa
The result is compared to previous measurements in Fig.~\ref{fig:f_lambdac_hera1}. The value obtained in DIS is consistent with a previous ZEUS measurement in the photoproduction regime~\cite{ref:f_lambdac_php} and with the $e^{+}e^{-}$ average value~\cite{ref:f_lambdac_ee}.

\section{Measurement of $D^{+}$ and $D^{0}$ cross sections with high precision and extraction of $F_{2}^{c\bar{c}}$}

The production of $D^{+}$ and $D^{0}$ mesons was measured in the kinematic region given by: $5 < Q^{2} < 1000$~GeV$^{2}$, $0.02 < y < 0.7$, $1.5 < p_{T}(D) < 15$~GeV and $|\eta(D) < 1.6|$. The measurement of $D^{+}$ mesons~\cite{ref:dplus_prel} used the data collected between 2005 and 2007 corresponding to a luminosity of $323 \pm 8$~pb$^{-1}$ while the $D^{0}$ measurement~\cite{ref:dplus_dzero_paper} was limited to $133.6 \pm 3.5$~pb$^{-1}$ of data collected in 2005. The decay channels $D^{+} \rightarrow K^{-}\pi^{+}\pi^{+}$ and $D^{0} \rightarrow K^{-}\pi^{+}$, and their charge conjugates, were reconstructed.

Lifetime information provided by the ZEUS microvertex detector~\cite{ref:mvd} was used to reduce the combinatorial background. This was achieved reconstructing displaced secondary vertices from $D$ meson decays.

\begin{figure}[h]
\centerline{\includegraphics[width=0.8\textwidth]{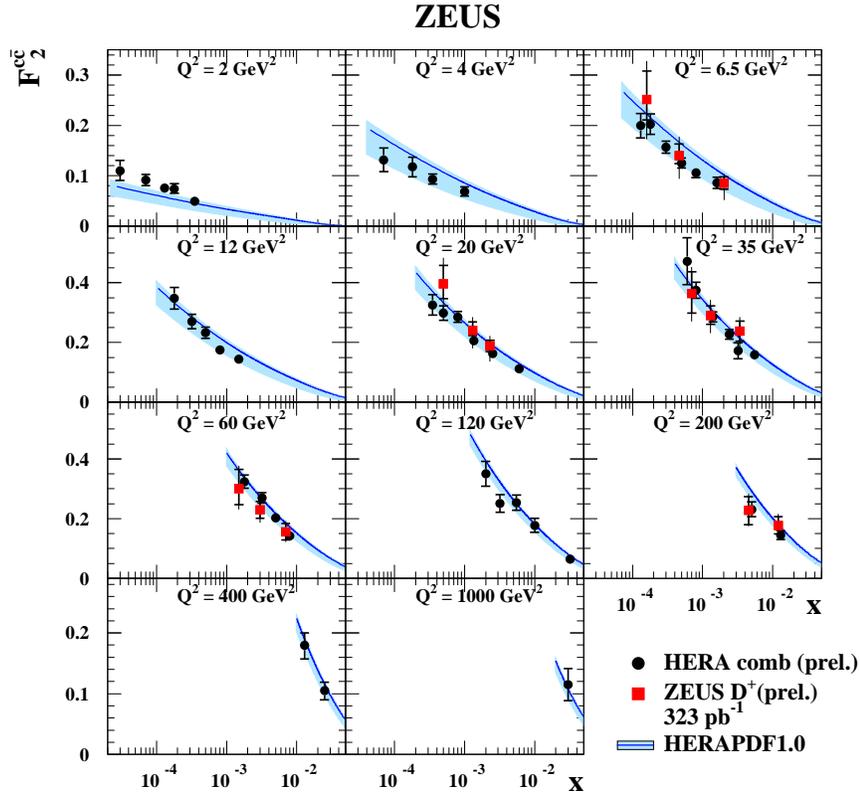}}
\caption{$F_{2}^{c\bar{c}}$ as a function of $x$ for fixed values of $Q^{2}$. Results obtained from $D^{+}$ decays are compared to H1 and ZEUS combined results and to the HERAPDF1.0 prediction.}
\label{fig:f2c_hera2}
\end{figure}

The double-differential cross section for $c\bar{c}$ production in DIS can be expressed as:
\beqa
\frac{d^{2} \sigma^{c\bar{c}}(x,Q^{2})}{dx dQ^{2}} = \frac{2 \pi \alpha^{2}}{Q^{4} x} [1 + (1 - y)^{2}] F^{c\bar{c}}(x,Q^{2}).
\eeqa
Here the small contribution to the cross section caused by $F_{L}^{c\bar{c}}$ was neglected. To extract $F_{2}^{c\bar{c}}$ from the measured double-differential cross sections in bins of $x$ and $Q^{2}$, an extrapolation to the full kinematic phase space was performed using HVQDIS:
\beqa
F_{2,{\rm meas}}^{c\bar{c}}(x_{i},Q_{i}^{2}) = \frac{\sigma_{{\rm meas},i}}{\sigma_{{\rm HVQDIS},i}} \times F_{2,{\rm HVQDIS}}^{c\bar{c}}(x_{i},Q^{2}_{i}).
\eeqa
The prediction for $F_{2}^{c\bar{c}}$ from HVQDIS, $F_{2,{\rm HVQDIS}}^{c\bar{c}}$, is multiplied by the ratio of the measured, $\sigma_{{\rm meas},i}$, to the predicted, $\sigma_{{\rm HVQDIS},i}$, visible $D$ meson cross sections in a given bin $i$. Results obtained using $D^{+}$ mesons are shown in Fig.~\ref{fig:f2c_hera2}. The data are in good agreement with combined $F_{2}^{c\bar{c}}$ values obtained from several previous H1 and ZEUS measurements~\cite{ref:hera_combined_f2c} including the $D^{0}$ cross sections measured using the data collected in 2005. Additionally, the data are compared to the HERAPDF1.0 prediction~\cite{ref:herapdf1}. The $F_{2}^{c\bar{c}}$ points measured from $D^{+}$ mesons, the H1 and ZEUS combination and the HERAPDF1.0 expectation are based on independent data sets.

\section{Summary}

The production of the charmed hadrons $D^{+}$, $D^{0}$ and $\Lambda_{c}^{+}$ in DIS has been studied in detail at HERA. All measurements are reasonably well described by NLO QCD predictions. For the first time the threshold region and $f(c \rightarrow \Lambda_{c}^{+})$ were investigated in DIS using new decay channels. Precise measurements of $D^{+}$ and $D^{0}$ mesons have been obtained using lifetime tags and the corresponding contribution to $F_{2}$ has been extracted.


\begin{thebibliography}{99}

\bibitem{ref:dplus_lambdac_paper} ZEUS Coll., H.~Abramowicz et al., \emph{Measurement of $D^{+}$ and $\Lambda_{c}^{+}$ production in deep inelastic scattering at HERA}, \emph{JHEP} {\bf 11}, 009 (2010).

\bibitem{ref:hvqdis} B.W.~Harris and J.~Smith, \emph{Charm quark and $D^{*\pm}$ cross sections in deeply inelastic scattering at DESY HERA}, \emph{Phys. Rev.} {\bf D 57}, 2806 (1998).

\bibitem{ref:zeus_s} ZEUS Coll., S.~Chekanov et al., \emph{ZEUS next-to-leading-order QCD analysis of data on deep inelastic scattering}, \emph{Phys. Rev.} {\bf D 67}, 012007 (2003).

\bibitem{ref:d_mesons_dis_zeus_hera1} ZEUS Coll., S.~Chekanov et al., \emph{Measurement of $D$ mesons production in deep inelastic scattering at HERA}, \emph{JHEP} {\bf 07}, 074 (2007).

\bibitem{ref:f_lambdac_php} ZEUS Coll., S.~Chekanov et al., \emph{Measurement of charm fragmentation ratios and fractions in photoproduction at HERA}, \emph{Eur. Phys. J.} {\bf C 44}, 351 (2005).

\bibitem{ref:f_lambdac_ee} L.~Gladilin, \emph{Charm Hadron Production Fractions}, {\tt hep-ex/9912064} (1999).

\bibitem{ref:dplus_prel} ZEUS Coll., \emph{Measurement of $D^{\pm}$ production and $F_{2}^{c}$ extraction in deep inelastic scattering at ZEUS}, ZEUS-prel-10-005 (2010).

\bibitem{ref:dplus_dzero_paper} ZEUS Coll., S.~Chekanov et al., \emph{Measurement of $D^{\pm}$ and $D^{0}$ production in deep inelastic scattering using a lifetime tag at HERA}, \emph{Eur. Phys. J.} {\bf C 63}, 171 (2009).

\bibitem{ref:mvd} A.~Polini et al., \emph{The design and performance of the ZEUS micro vertex detector}, \emph{Nucl. Inst. Meth.} {\bf A 581}, 656 (2007).

\bibitem{ref:hera_combined_f2c} H1 Coll. and ZEUS Coll., \emph{Combination of $F_{2}^{c\bar{c}}$ from DIS measurements at HERA}, H1prelim-09-171, ZEUS-prel-09-015 (2009).

\bibitem{ref:herapdf1} H1 Coll. and ZEUS Coll., F.D.~Aaron, \emph{Combined measurement and QCD analysis of the inclusive $e^{\pm}$ scattering cross sections at HERA}, \emph{JHEP} {\bf 01}, 109 (2010).

\end{thebibliography}
\end{document}